\documentclass{article}
\usepackage{arxiv}
\usepackage{graphicx}
\usepackage{mathtools}
\usepackage[utf8]{inputenc} 
\usepackage[T1]{fontenc}    
\usepackage{url}            
\usepackage{booktabs}       
\usepackage{amsfonts}       
\usepackage{nicefrac}       
\usepackage{microtype}      
\usepackage{lipsum}

\title{Investigation of Singing Voice Separation for Singing Voice Detection in Polyphonic Music}

\author{
  Yifu Sun \\
   School of Computer Science and Technology\\
  Fudan University\\
  Shanghai, 200438 China\\
  \texttt{yfsun20@fudan.edu.cn} \\
  \And
  Xulong Zhang\thanks{https://orcid.org/0000-0001-7005-992X} \\
  Ping An Technology (Shenzhen) Co., Ltd.\\
  Shenzhen, 518000 China\\
  \texttt{zhangxulong066@pingan.com.cn} \\
   \And
 Yi Yu \\
  Digital Content and Media Sciences Research Division\\ National Institute of Informatics\\
  Tokyo, 163-8001 Japan \\
  \texttt{yiyu@nii.ac.jp} \\
   \AND
   Xi Chen \\
   School of Computer Science and Technology\\
  Fudan University\\
  Shanghai, 200438 China\\
  \texttt{xichen1996@163.com} \\
   \And
   Wei Li \\
   $^1$ School of Computer Science and Technology \\
   $^2$ Shanghai Key Laboratory of Intelligent Information Processing\\
   Fudan University\\
   Shanghai, 200433 China\\
   \texttt{weili-fudan@fudan.edu.cn} \\
}

\begin{document}
\maketitle

\begin{abstract}
	Singing voice detection (SVD), to recognize vocal parts in the song, is an essential task in music information retrieval (MIR). The task remains challenging since singing voice varies and intertwines with the accompaniment music, especially for some complicated polyphonic music such as choral music recordings. To address this problem, we investigate singing voice detection while discarding the interference from the accompaniment. The proposed SVD has two steps: i. The singing voice separation (SVS) technique is first utilized to filter out the singing voice's potential part coarsely. ii. Upon the continuity of vocal in the time domain, Long-term Recurrent Convolutional Networks (LRCN) is used to learn compositional features. Moreover, to eliminate the outliers, we choose to use a median filter for time-domain smoothing. Experimental results show that the proposed method outperforms the existing state-of-the-art works on two public datasets, the Jamendo Corpus and the RWC pop dataset.
\end{abstract}


\section{Introduction}
\label{sec:introduction}

The purpose of singing voice detection task is to identify the part of the vocals in a given song. In the domain of music information retrieval (MIR), singing voice detection is often regarded as a pre-processing method to identify the vocal segments in the original mixed audio signal which can be exploited in many research topics, such as singer identification \cite{b22,zhang2021singer}, melody transcription \cite{b02}, query by humming \cite{b03}, lyrics transcription \cite{b04}, etc.  
	
	In the early years, researches had been focused on feature engineering. In \cite{rocamora2007comparing}, a feature set consisting of fluctogram, spectral flatness, spectral contraction and vocal variance, is chosen in order to distinguish the singing voice from highly harmonic instruments. But the identification result based on traditional techniques can hardly be considered ideal. As deep learning is successfully applied in feature representation and classification, better performance can be achieved. In \cite{kum2019joint}, Kum et al. used the multi-task learning method to detect and classify the singing voice jointly. In \cite{hou2020transfer}, better performance is achieved by transferring knowledge learned from a speech activity detection dataset.

\begin{figure}[htb]
  \centering
  \centerline{\includegraphics[scale=1.0, width=8cm]{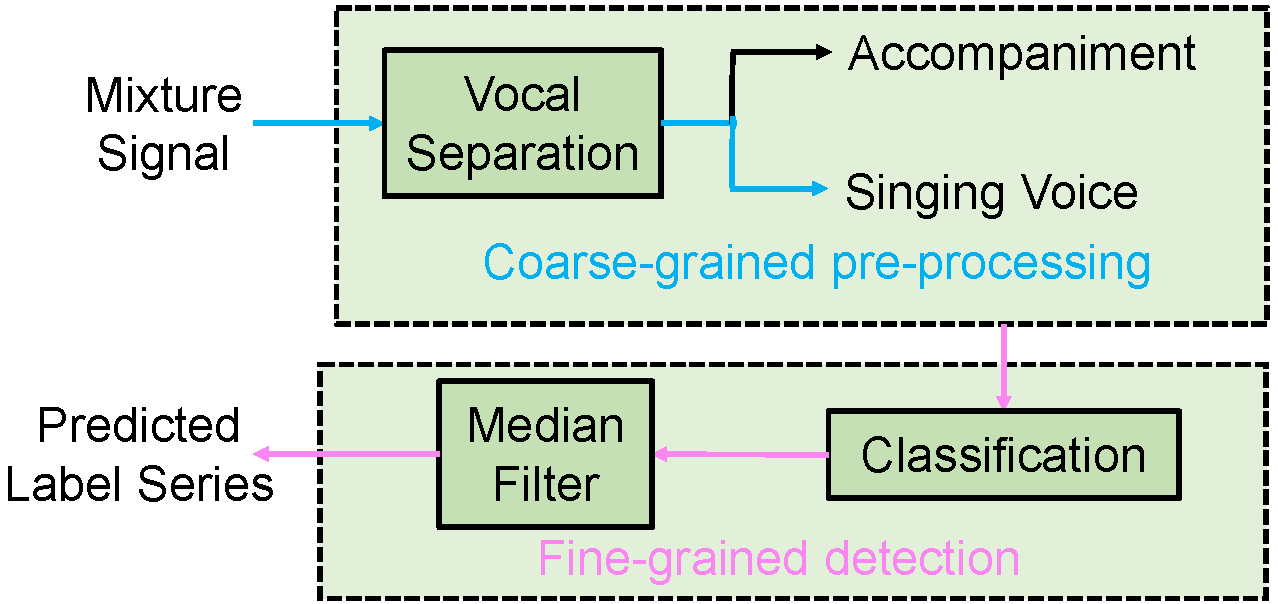}}
\caption{Idea of the SVD system with two steps}
\label{fig:SVDflowchart}
\end{figure}

Since singing voice varies through time and intertwines with the background accompaniment music, these factors lead to the difficulty of detecting singing voice activities in polyphonic music. The SVD task is similar to vocal detection (VD) in speech domain \cite{mauch2011timbre}. However, unlike random background noise, the accompaniment is highly correlated with the singing voice \cite{rafii2018overview}. The singing voice is a fluctuating sound, not as stationary as harmonic instruments like piano or guitar, but much more than percussive ones, like drums. It thus lies between harmonic and percussive components \cite{leglaive2015singing}.

According to the results of studying several SVS methods such as \cite{b13,b14,b15,b16,b17,b18} devised for analyzing non-stationary harmonic sound sources such as the singing voice, we propose to use U-Net \cite{jansson2017singing} to mitigate the interference of accompaniment. With the joint effort of the LRCN classifier \cite{donahue2015long} and the median filter as the post smoothing component \cite{lehner2013towards}, the idea of our SVD system containing two steps: coarse-grained pre-processing and fine-grained detection is proposed, as shown in Fig. \ref{fig:SVDflowchart}.

In our work, the four major procedures of vocal detection were compared, including vocal separation, feature selection, classifier selection, and filter smoothing. Finally we choose the process with the optimal performance in each step as our purposed method.

\section{Proposed Method}
\label{sec:ProposedMethod}

\begin{figure}[htb!]
  \centering
  \centerline{\includegraphics[scale=1.0, width=10cm]{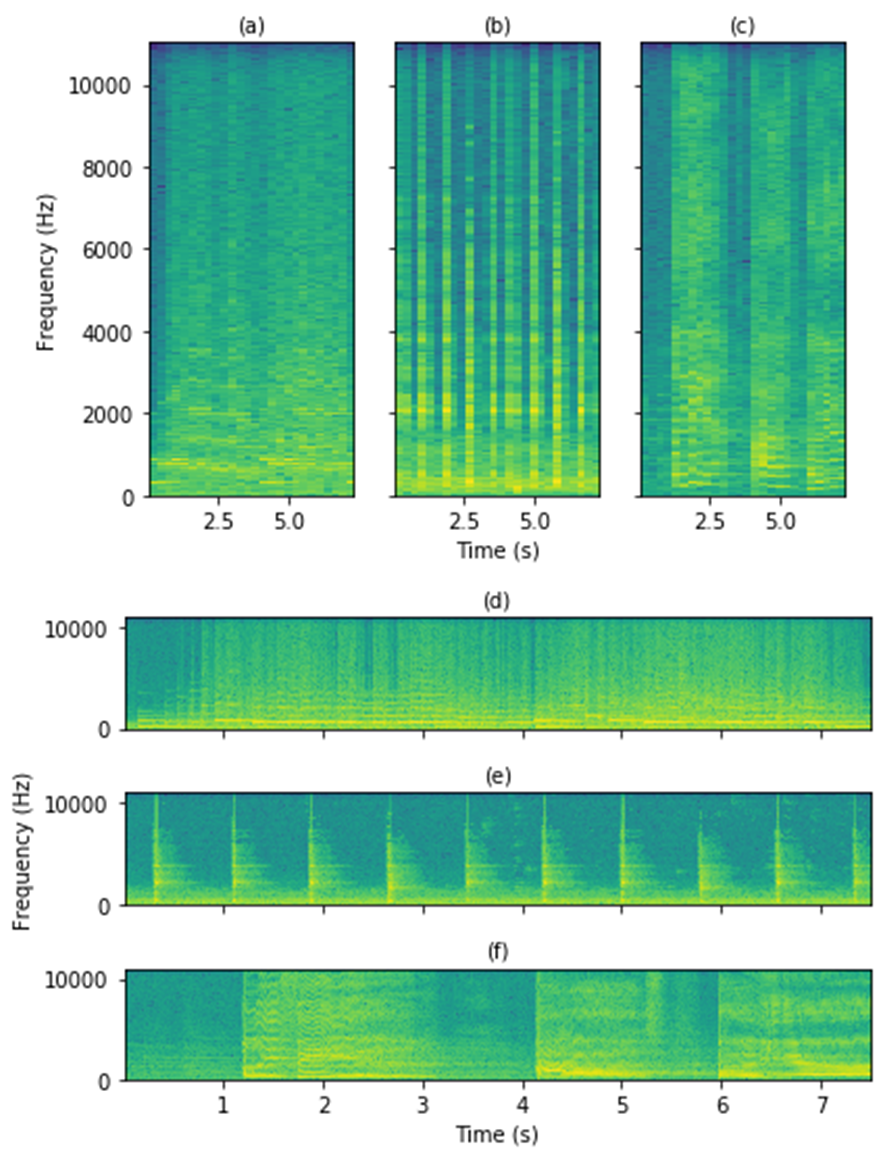}}
\caption{Long-window (186ms) spectrograms of (a), (b) and (c) corresponding to a piece of piano, drum and the vocal; Short-window (11ms) spectrograms of (d), (e) and (f) corresponding to a piece of piano, drum and the vocal. Above audio clips are in 44.1k Hz sample rate.  For long-window spectrograms, 8192 samples length, corresponding to 186ms, is used as the window size; For short-window, 512 samples length, corresponding to 11ms, is used as the window size.}
\label{fig:HPSSme2}
\end{figure}

Singing voice intertwining with the signal of background accompaniment makes SVD a challenging task. Fig. \ref{fig:HPSSme2} illustrates a detail comparison. The long-window spectrogram is used to concentrate more on the frequency domain for analysis. Since the singing voice does not have a relatively stable pitch like harmonic instruments, the energy spreads out a range of frequency channels (see Fig. \ref{fig:HPSSme2}(c)). This makes the singing voice acts more like a percussive instrument in Fig. \ref{fig:HPSSme2}(b), rather than a harmonic instrument in Fig. \ref{fig:HPSSme2}(a); under high time resolution, analysing using short-window spectrogram, pitch fluctuates of the vocal can be ignored, making energy concentrate on a set of frequency channels and smooth in the time direction. In this situation, the singing voice  (as seen Fig. \ref{fig:HPSSme2}(f)) is more like a harmonic instrument (as seen Fig. \ref{fig:HPSSme2}(d)).

In the audio processing field, it is typical to represent an audio clip in the time-frequency domain. Once using the mixture as input, no matter hand-crafted features or data-driven methods, the interference from time and frequency domain is hard to eliminate. This motivates us to propose using the SVS technique to remove the accompaniment before fine-grained singing voice detection.

\subsection{U-Net for singing voice separation}
\label{ssec:U-Net}

\begin{figure}[htb]
  \centering
  \centerline{\includegraphics[width=11cm]{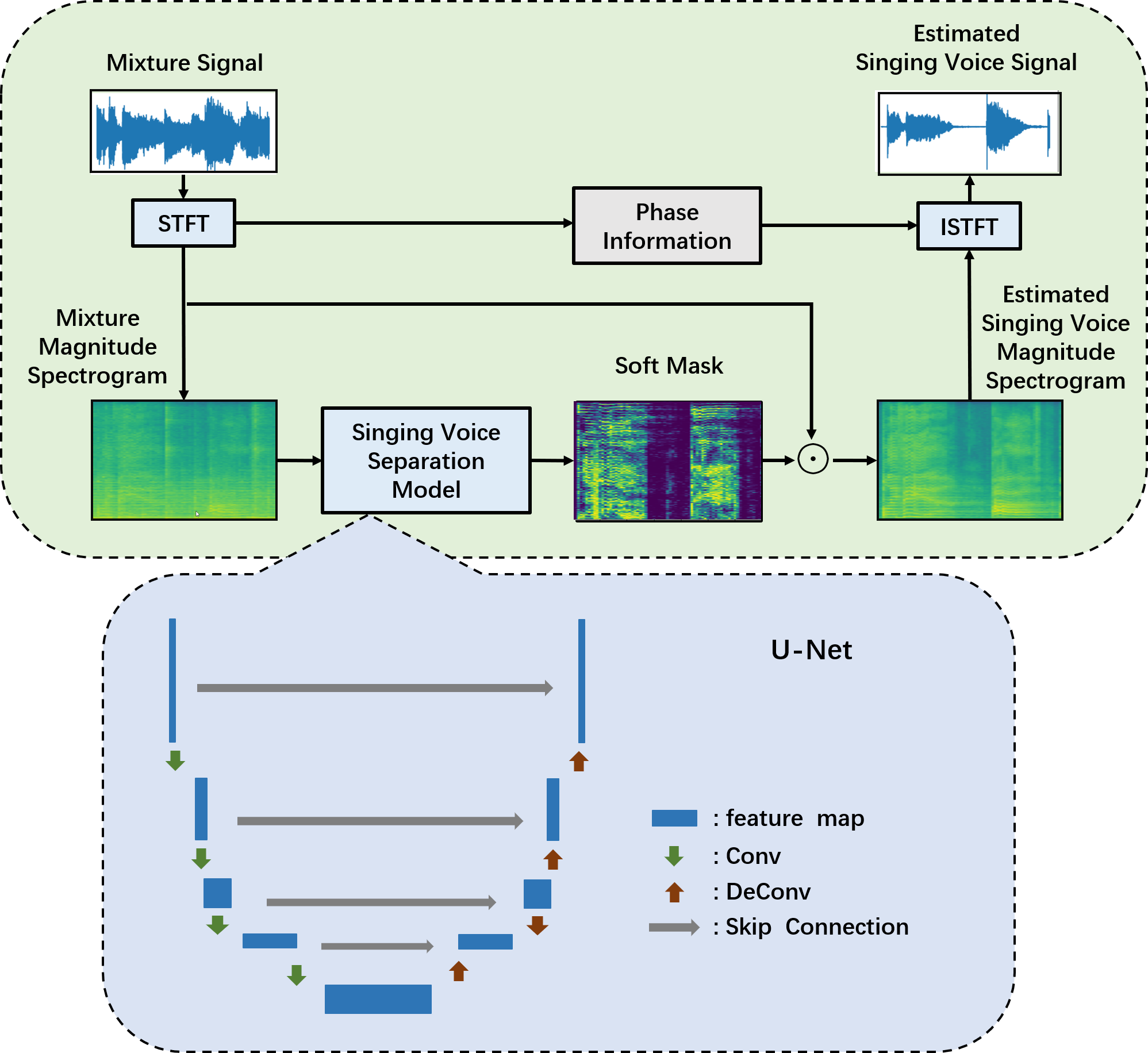}}
\caption{Singing voice separation method in the green dashed block and the U-Net model in the blue dashed block.}
\label{fig:UNet}
\end{figure}

The typical SVS method in the coarse-grained pre-processing step aims to generate the mask applied on the spectrogram to filter out the potential targeted part of the singing voice. How to model the target and generate the mask is the most challenging part. Over the years, SVS methods can be classified into three classes. The first assumpts that the singing voice is mostly harmonic \cite{rafii2018overview} and tries to model the singing voice by finding the fundamental frequency. According to the fact that the accompaniment is highly structed and tends to lie in a small range of frequencies \cite{rafii2018overview}, the second kind of method tries to model the accompaniment and get the singing voice by subtracting the accompaniment from the mixture. The third kind of method is the data-driven method which learns the model by large and representative datasets.

To recreate the fine, low-level detail required for high-quality audio reproduction \cite{ronneberger2015u}, the data-driven method, U-Net, is chosen as the SVS component for coarse-grained pre-processing. The flow chart of SVS using U-Net as the separator is shown in Fig. \ref{fig:UNet}. SVS's audio representation is the magnitude spectrogram obtained by short-time Fourier Transform (STFT). U-Net's target is to generate the mask used to extract the target spectrogram out \cite{fan2018svsgan}. In other words, the mask determines whether to keep constant or attenuated a certain frequency bin \cite{rafii2018overview}. Letting $ \odot $ denote the element wise manipulation, the masking operation is formulated as:

\begin{equation}
\hat{\textit{y}_{vocal}}\, =\, \textit{y}_{mix}\, \odot \, \textit{M},
\label{eq:maskOP}
\end{equation}

where $ \textit{y}_{mix}$, $ \hat{\textit{y}_{vocal}} $ and $ \textit{M} $ represent the spectrogram of the mixture, the estimated vocal spectrogram and the mask separately. Finally, the estimated vocal audio can be rebuilt using the phase information from STFT and inverse short-time Fourier Transform (ISTFT) operation. As illustrated in the dashed box in Fig. \ref{fig:UNet}, the encoder, the decoder and the skip connection build up the U-Net. A stack of convolutional operations composes the encoder. Each layer halves the feature map size in both dimensions simultaneously and doubles the number of channel, trying to encode smaller and deeper representations. In contrast, the decoder goes exactly the opposite way. The skip connections between the same hierarchical level allow low-level information to flow directly from the high-resolution input to the high-resolution output to recreate much more detailed information\cite{jansson2017singing}.

\subsection{Feature Extraction}
\label{sec:FeatureExtraction}

\begin{figure}[htb]
  \centering
  \centerline{\includegraphics[width=8.5cm]{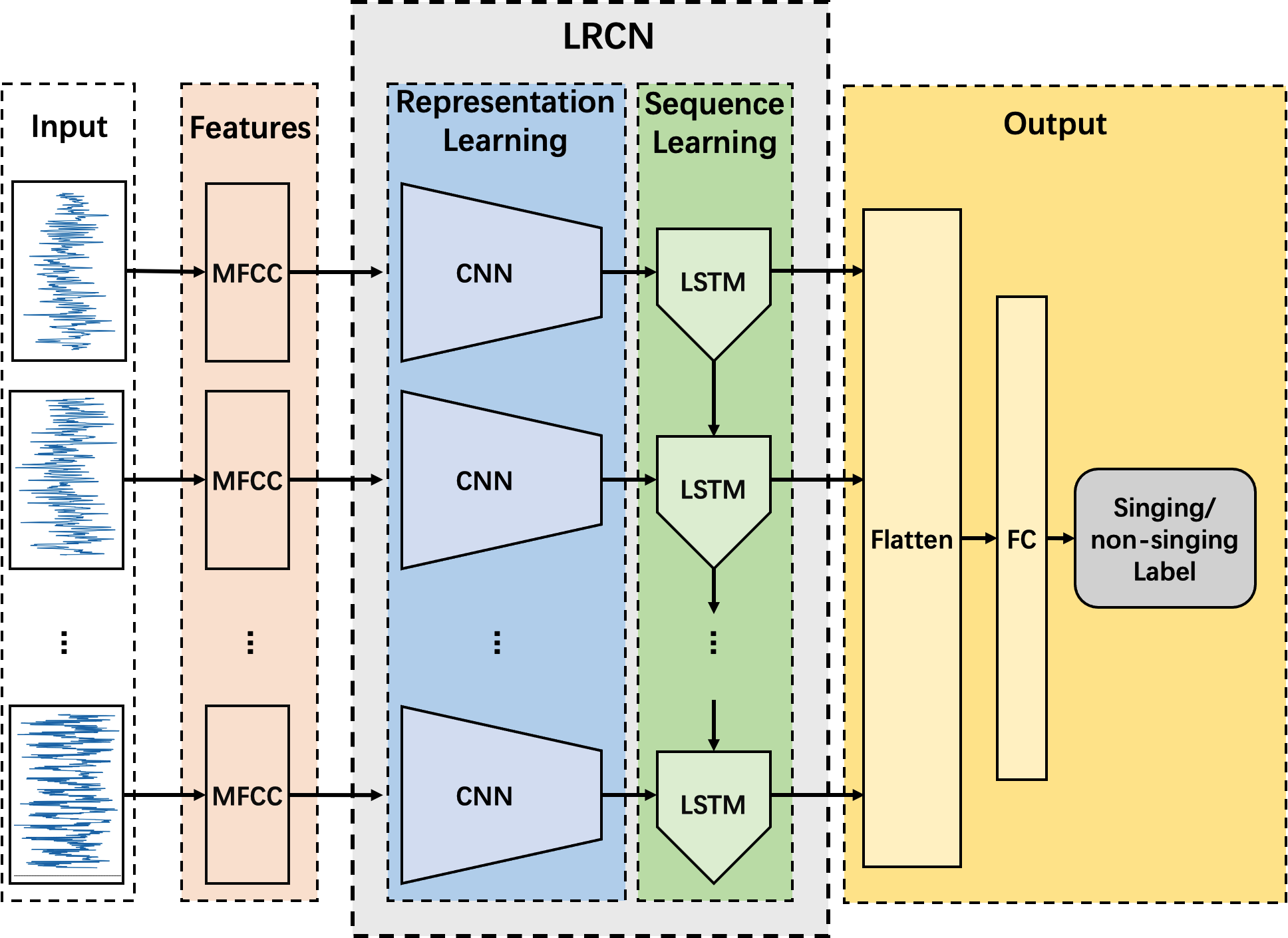}}
\caption{The LRCN model used for block-wise singing/non-singing classification.}
\label{fig:LRCN}
\end{figure}

There are many features proposed for the singing voice detection task. Among the features, most commonly used Mel-frequency Cepstral coefficients (MFCC) \cite{rocamora2007comparing}, Linear Predictive Cepstral Coefficients (LPCC) \cite{b24} and Perceptual Linear Predictive Coefficients (PLP) \cite{rocamora2007comparing} are chosen to examine the performance. MFCC has been widely used in a large number of speech and audio recognition tasks \cite{rocamora2007comparing}, and MFCC can represent the audio signal's timbre features. It is thus used as the feature in the proposed model. The most popular approach for modelling human voice production is Linear Prediction Coefficients (LPC) which performs well in a clean environment but not so good in a noisy one. LPCC are calculated by introducing the cepstrum coefficients in the LPC. Assume that LPCC, governed by the shape of vocal tract, is the nature of the sound. PLP is originally proposed by Hynek Hermansky as a way of warping spectra to minimize the differences between speakers while preserving important speech information.

The audio signal is first segmented into frames with overlapping. On each frame, a fast Fourier transform (FFT) is computed with a Hamming window. Most of the features are selected for their ability to discriminate voice from music \cite{rocamora2007comparing}.

Because MFCC can achieve the best performance among the other two features and their combinations, MFCC is chosen as the audio descriptor in our proposed singing voice detection method.

\subsection{LRCN for classification}
\label{sec:LRCNclassification}

Considering LRCN's capability of learning compositional acoustic representations in the feature and time domain, it is chosen as the classifier in the step of fine-grained singing voice detection. The LRCN architecture is a combination of Convolutional Neural Network (CNN) and Long Short-Term Memory (LSTM) which is shown in Fig. \ref{fig:LRCN}. And the key equations of LSTM structure are shown in Equation (\ref{eq:1})-(\ref{eq:5}).

\begin{equation}
	i(t) =\sigma(W_{i}\cdot[F_c(X(t)),H(t-1),C(t-1)] 
	  +b_{i}) 
\label{eq:1}
\end{equation}

\begin{equation}
	f(t)=\sigma(W_{f}\cdot[F_c(X(t)),H(t-1),C(t-1)]
	  + b_{f})
\label{eq:2}
\end{equation}

\begin{equation}
	\begin{split}
	C(t)={} &f(t)\cdot C(t-1)+i(t)\cdot \tanh(W_{c}\cdot  [F_c(X(t)), \\
	 & H(t-1)]+ b_{c})\\
\end{split}
\label{eq:3}
\end{equation}

\begin{equation}
	o(t)=\sigma(W_{o}\cdot [F_c(X(t)),H(t-1),C(t)] +b_{o}
\label{eq:4}
\end{equation}

\begin{equation}
	H(t)=o(t)\cdot \tanh{(C(t))}
	\label{eq:5}
\end{equation}

Where ‘$\cdot$’ represents the element-wise product, and ‘$F_c$’ represents the convolution operator. $\sigma$ is the sigmoid function, $W$ is the weight matrix, and $b$ is the bias vector. The input gate $i(t)$, forget gate $f(t)$ and output gate $o(t)$ of LRCN are separately listed in (\ref{eq:1}), (\ref{eq:2}) and (\ref{eq:4}). The $C(t)$ shown in Equation (\ref{eq:3}) is the LRCN cell, and $H(t)$ in Equation (\ref{eq:5}) is the output of the LRCN cell.

Since MFCC provides abundant information for SVD. Therefore, the CNN block in LRCN serves as the feature extractor for singing/non-singing information. The LSTM block in LRCN learns the long-range dependencies between different frames.

With drawbacks existing in the LSTM block of LRCN, the proposed system takes block-wise audio series as the input. Firstly, it is known that LSTM can solve the gradient vanishing problem to a certain extend. However, when it comes to long (such as more than 1000 steps, which is common in the audio processing field) series, the gradient vanishing problem is hard to eliminate in LSTM. Secondly, LSTM can not process information in parallel, making the training process time-consuming. Thirdly, the singing voice duration is often in word-level or sentence-level, while the dependency between frames is not that long. Therefore, block-wise data is processed as a unit in the LRCN classifier. In the last, with a flattened layer and three fully connected layers following the LRCN, the block-wise output is obtained.

\subsection{Post Smoothing}
\label{ssec:median filter}

Frame-wise classifiers tend to provide noisy results leading to an over-segmentation (short segments). Human annotation tends to provide under-segmentation (long segments ignoring instrumental breaks or singer breathing). For these reasons, post-processing was usually applied to the estimated segmentation. Accumulating the segment likelihood over a more extended period is more reliable for decision making. The median filter is applied as a post smoothing step. The label in each time frame is recalculated with a certain kind of the averaged value over a fixed size window along the time dimension.

\section{Evaluation}
\label{sec:experiment}

\subsection{Experiment Settings}
\label{ssec:settings}

The architecture of the pre-processing model, U-Net, is following that in \cite{jansson2017singing}. Each layer is composed of a 2D 5x5 kernel sized convolutional operation with stride 2, batch normalization, and leaky rectified linear units (ReLU) in the encoder leakiness 0.2. Each layer comprises a 2D 5x5 kernel sized deconvolutional operation with stride 2, batch normalization, and ReLU activation in the decoder. The dropout operation with a drop rate of 50\% is used in the first three layers.

The acoustic feature of the successive frames in a fixed duration block is used as the input for the classifier. The block duration is set to 600 ms, which is equal to 29 audio frames. The LRCN comprises a 2D 1x4 kernel sized convolutional layer with ReLu activation and an LSTM block with hard sigmoid activation. After flattening for each LSTM output, three fully connected layers with an output size of 200, 50 and 1 separately are added to get the final block-wise label.

\subsection{Dataset}
\label{ssec:dataset}
Two public datasets are used for a fair comparison of our approach with others. The Jamendo Corpus, retrieved from the Jamendo website, contains 93 copyright-free songs with singing voice activity annotations. The database is built and published along with \cite{ramona2008vocal}. The corpus is divided into three sets: the training set containing 61 songs, the validation set and the test set containing 16 songs, respectively. The RWC music dataset consists of 100 pop songs released by Goto \textit{et al.}. \cite{goto2002rwc}, while singing voice annotations are provided by Mauch \textit{et al}. \cite{mauch2011timbre}.

\subsection{Experiments and Results}
\label{ssec:ER}

\subsubsection{Singing voice separation as a pre-processing step}
\label{ssec:verifunet}

\begin{figure}[htb]
  \centering
  \centerline{\includegraphics[width=8.5cm]{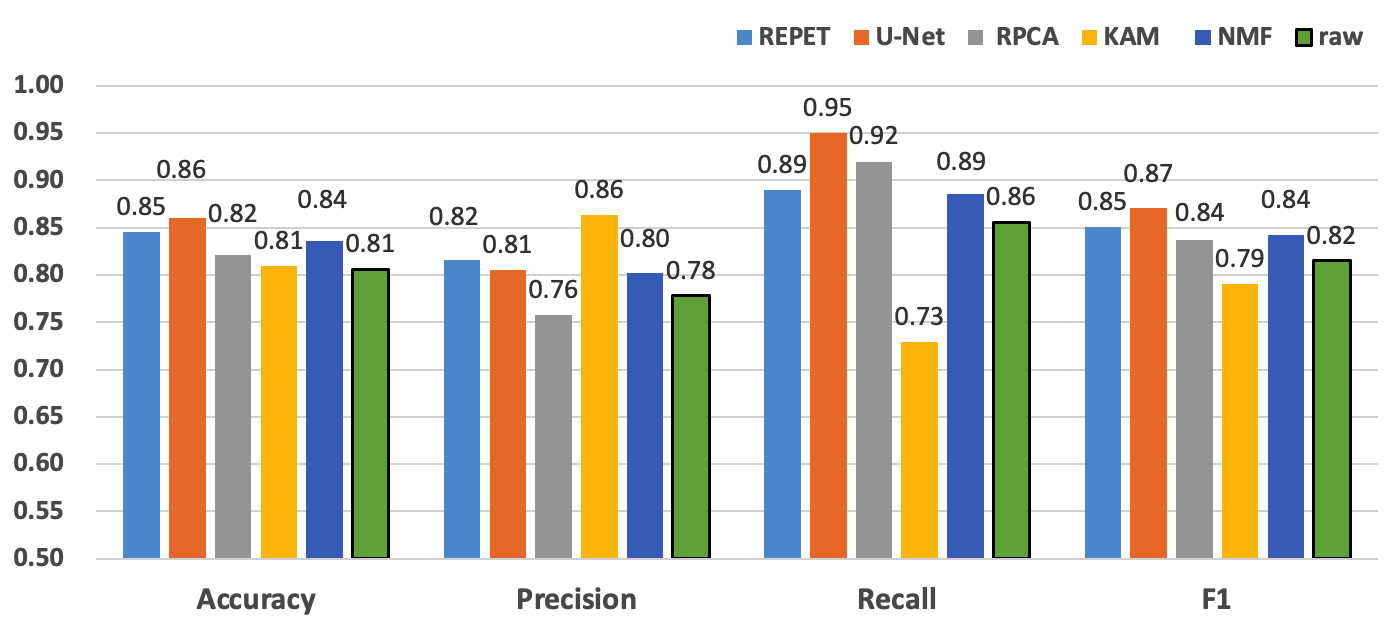}}
\caption{Comparison results on different SVS methods as the pre-processing step. The "raw" indicating the baseline method without pre-processing step.}
\label{fig:SVSCom}
\end{figure}

To demonstrate the importance of the singing voice separation method, especially the U-Net, as a pre-processing step on SVD, experiments are conducted on Jamendo Corpus in this section. We choose five methods as the pre-processing component, which are REpeating Pattern Extraction Technique (REPET) \cite{rafii2012repeating}, Kernel Additive Modelling (KAM) \cite{liutkus2014kernel}, Robust Principal Component Analysis (RPCA) \cite{huang2012singing}, multi-stage non-negative matrix factorization (NMF) \cite{zhu2013multi} from assumption-based methods and the U-Net\cite{donahue2015long}.

Experiment results can be seen in Fig. \ref{fig:SVSCom}. Measurements used are accuracy, precision, recall and F1. Except that the accuracy score of RPCA is slightly lower than that of the original method without pretreatment, the three methods (rept, U-Net and RPCA) are better than the original method in other measurements. Performances on recall and F1 of KAM are lower than that of the raw. This might be because that KAM does not only reduce the accompaniment but also affect the singing voice to some extent.

What is more, U-Net performs best as the pre-processing component. It is because data-driven methods do not have the assumptions that set a ceiling for the performance. The rationality of using U-Net as the pre-processing step is thus verified.

\subsubsection{post process of smoothing}

In this work, we evaluated two novelty detection approaches that used the same framework as the post-processing for the vocal detection system.

The first was a simple median filter with a fixed window length of 87 frames (3.48s) which was found to give the best trade-off between complexity and accuracy.

The second was Hidden Markov Model (HMM) \cite{b06} based method, a temporal smoothing of the posterior probabilities with a hidden Markov model that helped to adapt the segmentation sequence to the manual annotation.

Comparing the raw system framework without the post process, using the median filter as the temporal smoothing has improved in the f1 value, accuracy, and precision. With the comparison results, we chose the median filter for post-processing.

\subsubsection{Comparison results on the public dataset}
\label{ssec:withsota}

Finally, the proposed singing voice detection system is compared with the existing state-of-the-art works, i.e. Ramona \cite{ramona2008vocal}, Schlüter \cite{schluter2015exploring}, Lehner-1 \cite{lehner2014reduction}, Lehner-2 \cite{lehner2013towards}, Leglaive \cite{leglaive2015singing} and JDCs \cite{kum2019joint} on the Jamendo corpus, and with Mauch \cite{mauch2011timbre}, Schlüter \cite{schluter2015exploring}, Lehner-1 \cite{lehner2014reduction} and Lehner-2 \cite{lehner2013towards} on the RWC pop dataset. The comparison results are shown in the Table \ref{fig:J} and Table \ref{fig:R}. Our proposed system is named as U-Net-LRCN.

\begin{table}
	\centering
	\footnotesize
	\caption{Comparison results on Jamendo Corpus}
	\label{fig:J}
	\begin{tabular}{ccccc}
		\toprule
		Methods      & Accuracy & Precision & Recall & F1  \\
		\midrule
		Ramona \cite{ramona2008vocal}  & 0.822  & -     & -   & 0.831 \\
		Schlüter \cite{schluter2015exploring} & 0.923  & -     & 0.903 & -   \\
		Lehner-1 \cite{lehner2014reduction} & 0.882  & 0.880   & 0.862 & 0.871 \\
		Lehner-2 \cite{lehner2013towards} & 0.848  & -     & -   & 0.846 \\
		Leglaive \cite{leglaive2015singing} & 0.915  & 0.895   & 0.926 & 0.910 \\
		JDCs \cite{kum2019joint}& 0.800  & 0.791   & 0.802 & 0.792 \\
		U-Net-LRCN   & 0.888  & 0.865   & 0.920  & 0.892 \\
		\bottomrule
	\end{tabular}
\end{table}

\begin{table}
	\centering
	\footnotesize
	\caption{Comparison results on RWC pop dataset}
	\label{fig:R}
	\begin{tabular}{ccccc}
		\toprule
		Methods      & Accuracy & Precision & Recall & F1  \\
		\midrule
		Schlüter \cite{schluter2015exploring} & 0.927  & -     & 0.935 & -   \\
		Mauch \cite{mauch2011timbre}  & 0.872  & 0.887   & 0.921 & 0.904 \\
		Lehner-1 \cite{lehner2014reduction} & 0.875  & 0.875   & 0.926 & 0.900  \\
		Lehner-2 \cite{lehner2013towards} & 0.868  & 0.879   & 0.906 & 0.892 \\
		U-Net-LRCN        & 0.916  & 0.926   & 0.934 & 0.930 \\
		\bottomrule
	\end{tabular}
\end{table}

Table \ref{fig:J} shows comparison results on Jamendo Corpus. The U-Net-LRCN outperforms the shallow models. For the method of Leglaive \cite{leglaive2015singing}, the bi-LSTM model, which considers the past and the future information, is used. Leglaive \cite{leglaive2015singing} achieved an F1 of 0.91, which outperforms other state-of-the-art works. On F1, the U-Net-LRCN achieves 0.892, 0.018 lower than Leglaive \cite{leglaive2015singing}, is on par with state-of-the-art.

As seen in Table \ref{fig:R}, on RWC dataset, method of Schlüter \cite{schluter2015exploring} performs best. It uses a data argument method. The dataset used is not the original one. Besides, the precision and F1 score are not given. Except for Schlüter \cite{schluter2015exploring}, the U-Net-LRCN attained an F1 of 0.930, which is an improvement over the state-of-the-art method of Mauch \cite{mauch2011timbre} by 0.026.

In summary, the proposed U-Net-LRCN produces relatively better results compared with state-of-the-art methods over two public datasets. Using the SVS method as a pre-processing step has demonstrated its incredible power to eliminate the accompaniment's interference and improve SVD performance. However, our proposed U-Net-LRCN is also based on the LSTM to learn the context relation. The U-Net-LRCN is capable of learning the spatial relation and performs better than the LSTM.

\section{Conclusion}
\label{sec:conclu}

As singing voice changes and intertwines with the signal of background accompaniment in the time domain, it increases the difficulty of SVD in polyphonic music. Therefore, we propose to use the SVS method, U-Net, as a pre-processing step to eliminate the inference of background accompaniment. With the LRCN classifier's joint effort and the median post smoothing method, the proposed SVD system performs relatively better than the current state-of-the-art works.

Future work will try much more light-weight methods to eliminate the inference from the background accompaniment. Furthermore, using the proposed singing voice detection system presented in this paper to specific use cases such as singer identification will be attempted.

\section*{Acknowledgement}
This work was supported by National Key R\&D Program of China(2019YFC1711800), NSFC(62171138).

\bibliographystyle{IEEEtran}

\bibliography{mybib}

\end{document}